\documentclass{article}
\usepackage{spconf,amsmath,graphicx}
\usepackage{url}
\usepackage{capt-of}
\usepackage{bm}
\usepackage{tikz}
\usetikzlibrary{positioning, calc, fit, shapes.geometric, backgrounds, arrows.meta}
\usepackage{multirow} 
\usepackage{booktabs} 
\usepackage{standalone}
\usepackage{siunitx}
\usepackage{eso-pic}
\usepackage{stfloats}
\def\BibTeX{{\rm B\kern-.05em{\sc i\kern-.025em b}\kern-.08em
    T\kern-.1667em\lower.7ex\hbox{E}\kern-.125emX}}
\usepackage[acronym]{glossaries}
\newacronym{NVC}{NVC}{Neural Video Codec}
\newacronym{VCT}{VCT}{Video Compression Transformer}
\newacronym{RD}{RD}{Rate-Distortion}
\newacronym{CNN}{CNN}{Convolutional Neural Network}
\newacronym{MLP}{MLP}{Multi-Layer Perceptron}
\newacronym{BD}{BD}{Bjøntegaard Delta}
\newacronym{LRP}{LRP}{Latent Residual Prediction}
\newacronym{SWA}{SWA}{Sliding Window Attention}
\newacronym{P-SWA}{P-SWA}{Parallel Sliding Window Attention in Neural Video Coding}
\newacronym{VVC}{VVC}{Versatile Video Coding}
\newacronym{HEVC}{HEVC}{High Efficiency Video Coding}
\newacronym{WPP}{WPP}{Wavefront Parallel Processing}
\newacronym{MSE}{MSE}{Mean Squared Error}

\definecolor{convblue}{RGB}{189, 215, 238}
\definecolor{factorizedpurple}{RGB}{204, 192, 218}
\definecolor{containergray}{RGB}{230, 230, 230}
\definecolor{bitgray}{RGB}{128, 128, 128}

\title{Parallel Context Modeling for Sliding Window Attention in Neural Video Coding}
\name{Alexander Kopte and Andr\'e Kaup \thanks{The authors gratefully acknowledge that this work has been funded by the Deutsche Forschungsgemeinschaft (DFG, German Research Foundation) under project number 567512406.}}
\address{Chair of Multimedia Communications and Signal Processing\\Friedrich-Alexander University Erlangen-Nuremberg\\
Cauerstr. 7, 91058 Erlangen, Germany}
\begin{document}
\ninept

\AddToShipoutPictureBG*{%
  \AtPageUpperLeft{%
    \raisebox{-3\baselineskip}{
      \makebox[\paperwidth][c]{
        \parbox{0.9\textwidth}{\centering\footnotesize 
          © 2026 IEEE. Personal use of this material is permitted. Permission from IEEE must be obtained for all other uses, in any current or future media, including reprinting/republishing this material for advertising or promotional purposes, creating new collective works, for resale or redistribution to servers or lists, or reuse of any copyrighted component of this work in other works.
        }%
      }%
    }%
  }%
}
\maketitle

\begin{abstract}
Most neural video codecs rely on temporal conditioning, which makes them susceptible to error propagation over long sequences. While Transformer-based architectures like the \gls{VCT} offer a drift-free alternative, they suffer from high computational complexity and inferior \gls{RD} performance. The recent \gls{SWA} addresses these shortcomings by reducing complexity and enhancing \gls{RD} performance, yet it restricts decoding to a strictly sequential raster-scan order, creating a critical bottleneck in decoding latency. To resolve this, we propose \gls{P-SWA}, utilizing diagonal wavefronts to enable parallel decoding. By embedding a hyperprior and introducing an accumulator to fuse side information and local spatial context, our method increases decoding speed by \SI{36}{\percent} over the parallel \gls{VCT}. Simultaneously, it achieves Bjøntegaard Delta-rate savings of up to \SI{10.0}{\percent} for I-frames and \SI{7.1}{\percent} for P-frames over the SWA baseline. 
\end{abstract}

\glsresetall

\section{Introduction}
\label{sec:intro}

Learned video compression has achieved rate-distortion performance competitive with traditional codecs~\cite{jia_towards_2025}. Yet, most state-of-the-art \glspl{NVC} rely on explicit or implicit conditioning of the feature transform on temporal context. While beneficial for compression efficiency, this dependency introduces error propagation, where quantization and prediction errors accumulate over time. A drift-free alternative is the paradigm established by Mentzer \textit{et al.}~\cite{mentzer_vct_2022} with the \gls{VCT}. Although originally motivated by architectural simplicity, their approach models temporal correlations solely within the entropy model, leaving the feature transform unconditioned. While this eliminates drift, \gls{VCT} suffers from inferior \gls{RD} performance and high computational complexity due to the use of overlapping windows in the latent space.


Recently, \gls{SWA} was introduced in~\cite{kopte2025slidingwindowattentionlearned} to improve upon this, enabling free information propagation across the latent space. This significantly enhances \gls{RD} performance while reducing complexity. Despite these gains, \gls{SWA} relies on a strict raster-scan decoding order, preventing parallelization and resulting in prohibitive decoding latencies.

In this paper, we propose \gls{P-SWA} to resolve this bottleneck. We restructure the context model to support diagonal wavefront decoding, enabling the parallel prediction of multiple symbols per step. Specifically, we integrate a hyperprior within the transformer network and employ an accumulator to fuse its output with the previously decoded spatial context, enabling this parallel prediction scheme. Our experiments demonstrate that this approach increases decoding speed by \SI{36}{\percent} compared to \gls{VCT} while simultaneously improving \gls{RD} performance, achieving \gls{BD}-rate savings of up to \SI{10.0}{\percent} for I-frames and \SI{7.1}{\percent} for P-frames, effectively closing the gap to state-of-the-art methods for intra-coding.

\section{Related Work}
\label{sec:related}

Parallelization is a fundamental feature of traditional hybrid codecs. \gls{HEVC}~\cite{sullivan_overview_2012} and \gls{VVC}~\cite{bross_overview_2021} employ \gls{WPP}, which leverages the fact that probability estimation for entropy coding relies primarily on the immediate spatial neighborhood. This allows the decoding of a new row to start as soon as the relevant neighbors in the row above are available, creating a diagonal decoding wavefront. However, \gls{WPP} essentially operates as a single active wavefront, meaning the number of sequential steps remains proportional to the frame resolution. In contrast, our approach generalizes this concept by strictly enforcing a repeating local dependency pattern. This allows us to spawn a new independent wavefront every $s$ steps, enabling the processing of multiple spatially distant diagonals in the exact same step. Crucially, this decouples the number of sequential decoding steps from the frame resolution, reducing it to a fixed constant $s$.

Despite the competitive performance of recent \glspl{NVC}, high decoding latency remains a major barrier to practical adoption, with real-time capability only recently demonstrated by~\cite{jia_towards_2025}. The primary bottleneck is spatial autoregression, which enforces strict sequential dependencies. Minnen \textit{et al.}~\cite{minnen_channel-wise_2020} addressed this by substituting spatial dependencies with channel-wise autoregression, effectively limiting the number of serial steps to a fixed constant independent of resolution. However, ignoring spatial correlations limits coding efficiency. To combine spatial context with parallelization, Li \textit{et al.}~\cite{li_hybrid_2022} introduced checkerboard masking, later expanded in DCVC-DC~\cite{li_neural_2023} to utilize distinct spatial prediction patterns across four channel groups. While effective, their implementation relies on masked adapters that require a full network pass for each autoregressive step. This inherently couples computational complexity with the number of steps, necessitating a lighter backbone to maintain a fixed complexity budget. Similarly, Xiang \textit{et al.}~\cite{xiang_mimt_2022} employ a transformer to dynamically predict the next tokens based on masked input. While this scheme yields competitive \gls{RD} performance, it inherently links the total computational cost to the number of decoding steps. In contrast, our architecture manages autoregression entirely via internal attention masking. This decouples computational complexity from the prediction order, allowing for arbitrary mask patterns and step counts without affecting the model size or per-pixel cost.

Furthermore, the aforementioned conditional approaches suffer from error propagation, causing degradation over long sequences. While mitigation strategies like periodic context resets exist~\cite{li_neural_2024}, Mentzer \textit{et al.}~\cite{mentzer_vct_2022} proposed the \gls{VCT} to eliminate drift entirely by removing temporal conditioning from the feature transform. However, \gls{VCT} suffers from inferior \gls{RD} performance and high complexity due to redundant overlapping window computations. \gls{SWA}~\cite{kopte2025slidingwindowattentionlearned} improved upon this, employing efficient sliding window attention to enhance both performance and theoretical complexity. Nevertheless, \gls{SWA} retains a strictly sequential raster-scan prediction order, resulting in significant latency. Our work bridges this gap by introducing a parallelizable diagonal wavefront decoding scheme to the \gls{SWA} architecture, maintaining a large effective receptive field while achieving decoding speeds comparable to checkerboard-based methods.

\section{Proposed Method}
\label{sec:proposed}

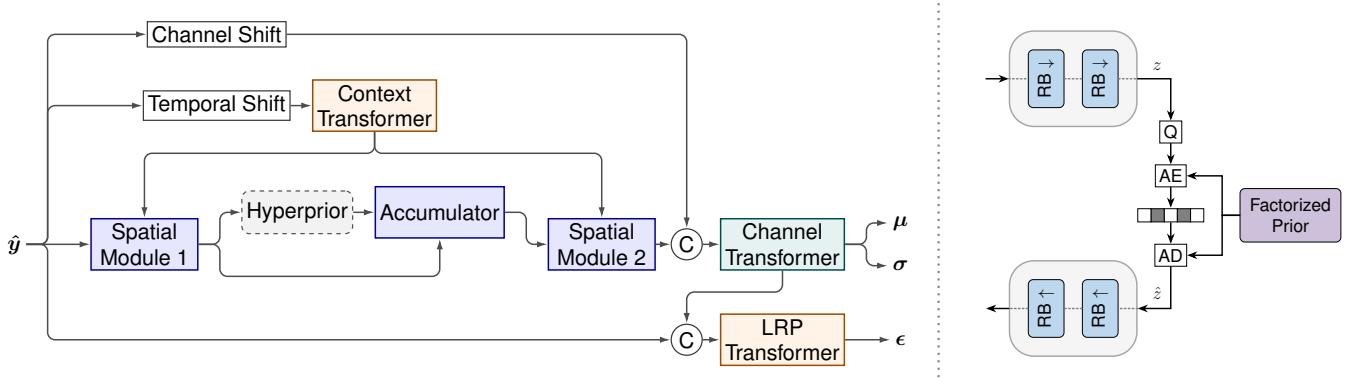
\begin{figure*}
    \centering
    \begin{minipage}[c]{0.69\textwidth}
        \begin{tikzpicture}[
    node distance=1.5cm and 2cm,
    font=\sffamily\Large,
    thick,
    base_block/.style={
        rectangle, 
        draw=black!80, 
        minimum width=2.5cm, 
        minimum height=1.2cm, 
        align=center,
    },
    tx_type1/.style={
        base_block,
        fill=blue!10,
        draw=blue!60!black,
        line width=1pt
    },
    tx_type2/.style={
        base_block,
        fill=orange!10,
        draw=orange!60!black,
        line width=1pt
    },
    tx_type3/.style={
        base_block,
        fill=teal!10,
        draw=teal!60!black,
        line width=1pt
    },
    general_block/.style={
        base_block,
        rounded corners=5pt,
        fill=gray!10,
        draw=gray!60!black,
        dashed
    },
    misc/.style={
        base_block,
        minimum width=0.5cm,
        minimum height=0.5cm,
        rounded corners=0pt,
    },
    connector/.style={
        ->,
        >={Latex[length=2.5mm, width=1.5mm]},
        draw=black!70,
        line width=1pt,
        rounded corners=5pt
    },
    bitstream/.style={
        connector,
        dashed
    }
]

    \node (input) {$\bm{\hat{y}}$};
    
    \node[tx_type1, right=1.5cm of input] (stt_pre) {Spatial\\Module 1};
    \node[general_block, right=1cm of stt_pre, yshift=0.75cm, minimum width=0.5cm, minimum height=1cm] (hyperprior) {Hyperprior};
    \node[tx_type1, right=0.5cm of hyperprior] (acc) {Accumulator};
    \node[tx_type1, right=1cm of acc, yshift=-0.75cm] (stt_post) {Spatial\\Module 2};
    \node[tx_type2] (tt) at ([shift={(0cm, 3.25cm)}] input -| acc.west) {Context\\Transformer};
    \node[tx_type3, right=1.5cm of stt_post] (ctt) {Channel\\Transformer};
    \node[tx_type2, below=1cm of ctt] (lrp) {LRP\\Transformer};
    \node[misc, left=0.4cm of lrp, circle, inner sep=2pt] (cat) {C};
    \node[misc, left=0.5cm of tt] (shift) {Temporal Shift};
    \node[misc, above=1cm of shift] (cshift) {Channel Shift}; 
    \node[misc, left=0.4cm of ctt, circle, inner sep=2pt] (ccat) {C};

    \node[right=1cm of ctt, yshift=0.5cm] (mean) {$\bm{\mu}$};
    \node[right=1cm of ctt, yshift=-0.5cm] (scale) {$\bm{\sigma}$};
    \node (lrp_) at (scale |- lrp) {$\bm{\epsilon}$};

    \draw[connector] (input.east) -- ++(0.5,0) |- (cshift.west);
    \draw[connector] (input.east) -- ++(0.5,0) |- (shift.west);
    \draw[connector] (input.east) -- ++(0.5,0) -- (stt_pre.west);
    \draw[connector] (input.east) -- ++(0.5,0) |- (cat.west);
    \draw[connector] (cshift.east) -| (ccat.north);
    \draw[connector] (shift.east) -- (tt.west);
    \draw[connector] (tt.south) -- ++(0,-0.5) -| (stt_pre.north);
    \draw[connector] (tt.south) -- ++(0,-0.5) -| (stt_post.north);
    \draw[connector] (stt_pre.east) -- ++(0.5,0) |- (hyperprior.west);
    \draw[connector] (stt_pre.east) -- ++(0.5,0) |- ++(1.5, -0.75) -| (acc.south);
    \draw[connector] (hyperprior.east) -- (acc.west);
    \draw[connector] (acc.east) -- ++(0.5,0) |- (stt_post.west);
    \draw[connector] (stt_post.east) -- (ccat.west);
    \draw[connector] (ccat.east) -- (ctt.west);
    \draw[connector] (ctt.south) -- ++(0,-0.5) -| (cat.north);
    \draw[connector] (cat.east) -- (lrp.west);

    \draw[connector] (ctt.east) -- ++(0.5,0) |- (mean.west);
    \draw[connector] (ctt.east) -- ++(0.5,0) |- (scale.west);
    \draw[connector] (lrp.east) -- (lrp_.west);

\end{tikzpicture}
    \end{minipage}
    \begin{tikzpicture}[baseline=0]
        \draw[dotted, thick, gray] (0,-2.35) -- (0, 2.6cm);
    \end{tikzpicture}
    \hfill
    \begin{minipage}[c]{0.27\textwidth}
        \begin{tikzpicture}[
    node distance=1.0cm and 1.0cm,
    font=\sffamily\Large,
    thick,
    block_base/.style={
        draw,
        rounded corners=4pt,
        align=center,       
        anchor=north,       
        minimum height=2.5em,
        inner sep=8pt
    },
    rb_block/.style={block_base, rotate=90, fill=convblue, minimum width=3.5em},
    prior/.style={block_base, fill=factorizedpurple, minimum width=3.5em},
    small_sq/.style={
        draw,
        fill=white,
        minimum size=2em,
        align=center
    },
    bit_sq/.style={draw, minimum size=1.2em}, 
    container_box/.style={
        draw=gray!70, fill=containergray!40, very thick, 
        rounded corners=20pt, inner sep=15pt
    },
    flow_arrow/.style={-Stealth, very thick, line cap=round},
    dotted_connect/.style={dotted, very thick, gray, line cap=round}
]


\node[rb_block] (rb1_down) at (0,0) {$\text{RB}\downarrow$};

\node[rb_block, right=0.5cm of rb1_down.south, anchor=north] (rb2_down) {$\text{RB}\downarrow$};

\draw[dotted_connect] (rb1_down) -- (rb2_down);


\coordinate (turn_point_1) at ($(rb2_down.south) + (1.5cm, 0)$);
\node[font=\Large, above left=0.2cm of turn_point_1] (z_label) {$z$};

\node[small_sq, below=1.2cm of turn_point_1] (Q) {Q};
\node[small_sq, below=0.6cm of Q] (AE) {AE};

\node[bit_sq, fill=white, below=0.6cm of AE] (b3) {};

\node[bit_sq, fill=bitgray, left=-\pgflinewidth of b3] (b2) {};
\node[bit_sq, fill=white, left=-\pgflinewidth of b2] (b1) {};
\node[bit_sq, fill=bitgray, right=-\pgflinewidth of b3] (b4) {};
\node[bit_sq, fill=white, right=-\pgflinewidth of b4] (b5) {};

\node[fit=(b1)(b5), inner sep=0pt] (bitstream) {};

\node[prior, right=1.0cm of bitstream] (prior) {Factorized\\Prior};

\node[small_sq, below=0.6cm of bitstream] (AD) {AD};


\coordinate[below=1.2cm of AD] (turn_point_2);
\node[font=\Large, above left=0.2cm of turn_point_2] (zhat_label) {$\hat{z}$};

\node[rb_block, anchor=center] (rb1_up) at (rb2_down.center |- turn_point_2) {$\text{RB}\uparrow$};
\node[rb_block, anchor=center] (rb2_up) at (rb1_down.center |- turn_point_2) {$\text{RB}\uparrow$};

\draw[dotted_connect] (rb1_up) -- (rb2_up);


\begin{pgfonlayer}{background}
    \node[container_box, fit=(rb1_down) (rb2_down)] (top_container) {};
    \node[container_box, fit=(rb1_up) (rb2_up)] (bottom_container) {};
\end{pgfonlayer}

\draw[flow_arrow] ($(rb1_down.north) + (-1.2cm, 0)$) -- (top_container.west);
\draw[dotted_connect] (top_container.west) -- (rb1_down.north);

\draw[dotted_connect] (rb2_down.south) -- (top_container.east);
\draw[flow_arrow] (top_container.east) -- (turn_point_1) -- (Q.north);

\draw[flow_arrow] (Q) -- (AE);
\draw[flow_arrow] (AE) -- (bitstream);
\draw[flow_arrow] (bitstream) -- (AD);
\draw[flow_arrow] (prior.west) -- ++(-0.5, 0) |- (AE);
\draw[flow_arrow] (prior.west) -- ++(-0.5, 0) |- (AD);

\draw[flow_arrow] (AD.south) |- (bottom_container.east);
\draw[dotted_connect] (bottom_container.east) -- (rb1_up.south);

\draw[dotted_connect] (rb2_up.north) -- (bottom_container.west);
\draw[flow_arrow] (bottom_container.west) -- ($(rb2_up.north) + (-1.2cm, 0)$);

\end{tikzpicture}
    \end{minipage}
    \caption{Overview of the proposed \gls{P-SWA} architecture (\textbf{left}) and the hyperprior network (\textbf{right}). We denote channel concatenation (C), quantization (Q), and arithmetic coding (AE/AD). $\text{RB}\downarrow$ and $\text{RB}\uparrow$ are down- and upsampling residual blocks from DCVC-DC~\cite{li_neural_2023}.}
    \label{fig:overview}
\end{figure*}

\subsection{Overview}
The overall architecture of the proposed entropy model is illustrated in Fig.~\ref{fig:overview}. First, a \emph{Context Transformer} extracts temporal context from the quantized latent $\bm{\hat{y}}$ using time-causal \gls{SWA}~\cite{kopte2025slidingwindowattentionlearned}. Briefly, \gls{SWA} restricts standard self-attention to a localized 3D sliding window, significantly reducing computational complexity while preserving local spatio-temporal dependencies. To ensure autoregressivity, a temporal shift aligns the previous frame's latent with the current frame's position, while a learned padding vector provides the initial context for the very first frame. Spatial processing is then handled by 2D \gls{SWA} masked for parallel decoding (Sec.~\ref{sec:parallel_decoding}). This is divided into two \emph{Spatial Modules} separated by an accumulator. Both modules alternate between spatial self-attention and temporal cross-attention. We revert to this separation because unifying them into a single attention mechanism~\cite{kopte2025slidingwindowattentionlearned} exhibited higher optimization variance, resulting in frequent training instabilities in our setup.

\emph{Spatial Module 1} processes the current frame conditioned on the temporal context. Its output feeds into a lightweight convolutional hyperprior to transmit lossy side information. We then introduce an accumulator to resolve the initialization problem inherent to parallel sliding window schemes: while standard approaches require zero-padding at the start of each wavefront, which disrupts spatial structures within the frame, our accumulator eliminates this need. Architecturally, it is implemented as a standard cross-attention block. The decoded hyperprior serves as the Query ($Q$) and the residual skip connection. To fetch lossless local context from already-decoded neighbors, the Key ($K$) and Value ($V$) are derived from the \emph{Spatial Module 1} output and masked to restrict attention strictly to past positions, as visualized in Fig.~\ref{fig:decoding_order}b. This cleanly initializes the diagonal wavefronts by fusing side information with lossless local context from already transmitted neighbors. Following~\cite{jia_towards_2025}, we also employ a filterbank for the hyperprior's latent space, utilizing distinct priors for each rate point and separate priors for the first four frames.

After processing by \emph{Spatial Module 2}, the latents enter a \emph{Channel Transformer} for channel-wise autoregression across $N$ channel groups. While a unified 4D spatio-temporal-channel attention would be ideal, preliminary experiments showed that P-frame prediction requires a high model dimension per token, making 4D attention computationally infeasible. Instead, we constrain channel interactions to a lightweight mixing layer within the \emph{Channel Transformer}. This layer replaces standard \gls{SWA} with a dense linear projection acting solely on the channel dimension. A block-lower-triangular mask applied to the weight matrix ensures causality: group $i$ only accesses groups $0 \dots i$. This acts as a masked $1 \times 1$ convolution, capturing inter-channel dependencies while preserving spatial token independence. A channel shift operation ensures the first channel group attends only to the spatio-temporal context.

Finally, two prediction heads, consisting of grouped two-layer \glspl{MLP}, predict the mean $\bm{\mu}$ and scale $\bm{\sigma}$ of the Gaussian distribution. The \gls{LRP}~\cite{minnen_channel-wise_2020} corrects the quantization error. Because it is applied after context modeling, it is spatially non-causal but must remain temporally causal. We therefore concatenate the final representation from the \emph{Channel Transformer} with the quantized latent $\bm{\hat{y}}$ and process it via a separate \emph{LRP Transformer} to predict the residual $\bm{\epsilon}$.

Furthermore, we modernize the transformer architecture by adopting \mbox{SwiGLU} activations and \mbox{RMSNorm} pre-normalization, following the Llama 2 architecture~\cite{touvron_llama_2023}. To support multiple rate points with a single network, we follow the strategy from~\cite{kopte2025slidingwindowattentionlearned}: since the feature transform controls rate via latent scaling, which \mbox{RMSNorm} subsequently removes, we introduce learnable channel-wise scaling parameters at the transformer's input and outputs to explicitly re-inject this rate information.

\subsection{Parallel decoding} \label{sec:parallel_decoding}

\begin{figure}[t]
    \centering
    \begin{minipage}[b]{0.48\columnwidth}
        \centering
        \begin{tikzpicture}[
    x=1cm, y=1cm,
    numstyle/.style={font=\sffamily\Large\bfseries},
    color1/.style={fill=red!30},
    color2/.style={fill=blue!30},
    color3/.style={fill=green!30},
    color4/.style={fill=orange!30},
    dotted-grid/.style={
        black, line width=2.5pt, line cap=round, 
        dash pattern=on 0pt off 5.69pt, 
        dash phase=0pt
    }
]

    \def\xmin{-2}
    \def\xmax{5}
    \def\ymin{-2}
    \def\ymax{5}

    \foreach \row in {\ymin,...,\ymax} {
        \foreach \col in {\xmin,...,\xmax} {
            \pgfmathsetmacro{\op}{(\row >= 0 && \row <= 3 && \col >= 0 && \col <= 3) ? 1.0 : 0.3}
            \pgfmathparse{int(mod(mod(\row,4)+4,4))} \let\modRow\pgfmathresult
            \pgfmathparse{int(mod(mod(\col,4)+4,4))} \let\modCol\pgfmathresult
            \pgfmathparse{int(mod(\modRow + \modCol, 4) + 1)} \let\val\pgfmathresult

            \begin{scope}[shift={(\col, -\row)}]
                \fill[color\val, opacity=\op] (0,0) rectangle (1,-1);
                \draw[gray, thin, opacity=\op] (0,0) rectangle (1,-1);
                \pgfmathsetmacro{\textop}{\op < 1 ? 0.5 : 1}
                \node[numstyle, opacity=\textop] at (0.5, -0.5) {\val};
            \end{scope}
        }
    }

    \begin{scope}
        \clip (\xmin-0.1, -\ymin+0.1) rectangle (\xmax+1.1, -\ymax-1.1);
        
        \draw[dotted-grid] 
            ({floor(\xmin/4)*4}, {floor((-\ymax-1)/4)*4}) 
            grid[step=4] 
            ({ceil((\xmax+1)/4)*4}, {ceil(-\ymin/4)*4 + 4});
    \end{scope}

    \draw[black, line width=2.5pt, line join=miter] (0, 0) rectangle (4, -4);

\end{tikzpicture}
        \vspace{2pt}
        \centerline{(a)}
    \end{minipage}
    \hfill
    \begin{minipage}[b]{0.48\columnwidth}
        \centering
        \begin{tikzpicture}[
    x=1cm, y=1cm,
    numstyle/.style={font=\sffamily\Large\bfseries},
    color1/.style={fill=red!30},
    color2/.style={fill=blue!30},
    color3/.style={fill=green!30},
    color4/.style={fill=orange!30}
]

    \def\xmin{-2}
    \def\xmax{5}
    \def\ymin{-2}
    \def\ymax{5}
    
    \def\targetRow{1}
    \def\targetCol{1}
    \def\targetVal{3} 

    \pgfmathtruncatemacro{\wminRow}{\targetRow-2}
    \pgfmathtruncatemacro{\wmaxRow}{\targetRow+2}
    \pgfmathtruncatemacro{\wminCol}{\targetCol-2}
    \pgfmathtruncatemacro{\wmaxCol}{\targetCol+2}

    \foreach \row in {\ymin,...,\ymax} {
        \foreach \col in {\xmin,...,\xmax} {
            \pgfmathparse{int(mod(mod(\row,4)+4,4))} \let\modRow\pgfmathresult
            \pgfmathparse{int(mod(mod(\col,4)+4,4))} \let\modCol\pgfmathresult
            \pgfmathparse{int(mod(\modRow + \modCol, 4) + 1)} \let\val\pgfmathresult

            \pgfmathtruncatemacro{\inWindow}{
                (\col >= \wminCol && \col <= \wmaxCol && 
                 \row >= \wminRow && \row <= \wmaxRow) ? 1 : 0
            }
            
            \begin{scope}[shift={(\col, -\row)}]
                \def\cellOpacity{0.2}
                \def\txtOpacity{0.2}
                
                \ifnum\inWindow=1
                    \ifnum\val<\targetVal
                        \def\cellOpacity{1.0}
                        \def\txtOpacity{1.0}
                    \else
                        \ifnum\val=\targetVal
                            \def\cellOpacity{0.2}
                            \def\txtOpacity{0.2}
                        \fi
                    \fi
                \fi

                \fill[color\val, opacity=\cellOpacity] (0,0) rectangle (1,-1);
                \node[numstyle, opacity=\txtOpacity] at (0.5, -0.5) {\val};
                
                \def\drawGray{0}
                \ifnum\inWindow=0
                    \def\drawGray{1}
                \else
                    \ifnum\val=\targetVal
                        \def\drawGray{0} 
                    \else
                        \def\drawGray{1} 
                    \fi
                \fi
                
                \ifnum\drawGray=1
                    \draw[gray, thin, opacity=1.0] (0,0) rectangle (1,-1);
                \fi
            \end{scope}
        }
    }

    \foreach \row in {\wminRow,...,\wmaxRow} {
        \foreach \col in {\wminCol,...,\wmaxCol} {
            \pgfmathparse{int(mod(mod(\row,4)+4,4))} \let\modRow\pgfmathresult
            \pgfmathparse{int(mod(mod(\col,4)+4,4))} \let\modCol\pgfmathresult
            \pgfmathparse{int(mod(\modRow + \modCol, 4) + 1)} \let\val\pgfmathresult

            \begin{scope}[shift={(\col, -\row)}]
                \ifnum\val=\targetVal
                    \def\isTarget{0}
                    \ifnum\row=\targetRow
                        \ifnum\col=\targetCol
                            \def\isTarget{1}
                        \fi
                    \fi
                    \ifnum\isTarget=0
                        \draw[green!60!black, line width=2pt] (0,0) rectangle (1,-1);
                    \fi
                \fi
            \end{scope}
        }
    }

    \draw[black, line width=3.5pt, rounded corners=2pt] 
        (\targetCol - 2, -\targetRow + 2) rectangle (\targetCol + 3, -\targetRow - 3);

    \begin{scope}[shift={(\targetCol, -\targetRow)}]
        \draw[green!60!black, line width=3.5pt] (0,0) rectangle (1,-1);
    \end{scope}

\end{tikzpicture}
        \vspace{2pt}
        \centerline{(b)}
    \end{minipage}
    \caption{Parallel decoding scheme. (a) A repeating spatial mask creates diagonal wavefronts. (b) Receptive field within a \numproduct{5x5} window for a target hyperpixel (3, thick green border). Past hyperpixels (opaque) are visible to the accumulator. Other current-step hyperpixels (3, thin borders) are masked in the accumulator but unmasked for spatial self-attention. Future hyperpixels (4) remain masked.}
    \label{fig:decoding_order}
\end{figure}
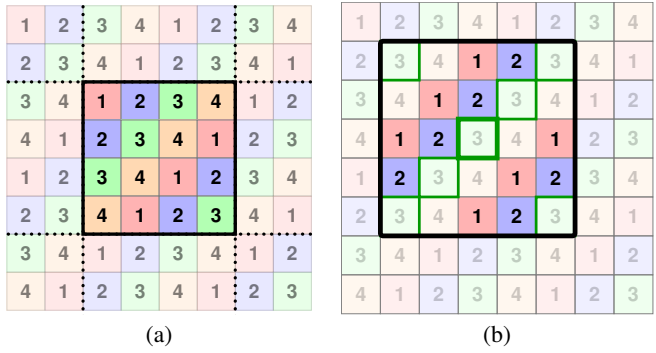

To achieve parallel decoding, we adapt the \gls{SWA} mechanism to utilize a repeating $s \times s$ spatial attention mask. This is realized by implementing a modified \gls{SWA} Triton~\cite{triton} kernel, which builds upon FlashAttention-2~\cite{dao_flashattention-2_2023}, to support arbitrary spatial masking. We adopt a specific pattern where each row of the mask is a cyclic shift of the previous row, as visualized in Fig.~\ref{fig:decoding_order}a. This creates a diagonal wavefront where every $s$-th diagonal is predicted simultaneously. 

Unlike I-frames, which rely on local texture features that are relatively agnostic to prediction direction, P-frames heavily depend on temporal context. A consistent prediction direction (from top-left to bottom-right) is crucial for the model to effectively learn implicit motion compensation within the latent space. Random or non-directional masks, such as the four-step pattern in~\cite{li_neural_2023}, disrupt this consistency and degrade P-frame performance.

In contrast to DCVC-DC~\cite{li_neural_2023}, which interleaves spatial and channel steps, we decouple them into $s$ spatial steps and $N$ channel groups. While interleaving reduces the total step count, it treats channel groups as distinct tokens, increasing sequence length by a factor of $N$. This would require reducing the model dimension $d$ for feasibility, severely degrading P-frame performance. Instead, decoupling restricts the spatial transformer to $s$ passes, handling channel interactions via a lightweight mixing layer. This preserves the high model capacity required for P-frames while keeping computational costs manageable.

\section{Experiments}

\begin{figure*}[!t]
    \centering
    \begin{minipage}{\textwidth}
        \centering
        \addtocounter{table}{1}
        \vspace{-5pt}
        \captionof{table}{\gls{BD}-rate savings (\%) relative to the SWA baseline~\cite{kopte2025slidingwindowattentionlearned} for all test datasets.}
        \label{experiments:results:table}
        \vspace{0.2cm}

        \sisetup{
            retain-explicit-plus,
            table-format = +-2.1,
            detect-weight = true, 
            mode = text 
          }
        \begin{tabular}{l l S S S}
            \toprule
            \textbf{Dataset} & \textbf{Coder} & \textbf{BD-Rate (I-frames)} & \textbf{BD-Rate (P-frames)} & \textbf{BD-Rate (GOP)} \\
            \midrule
            \multirow{6}{*}{UVG} 
            & P-SWA (ours) & -9.6\,\si{\percent} & -4.6\,\si{\percent} & -5.0\,\si{\percent} \\
            & VCT~\cite{mentzer_vct_2022} & +28.2\,\si{\percent} & +20.7\,\si{\percent} & +21.5\,\si{\percent} \\
            & DCVC-DC~\cite{li_neural_2023} & -11.8\,\si{\percent} & -49.6\,\si{\percent} & -45.0\,\si{\percent} \\
            & HM 18.0~\cite{HM18} & +36.7\,\si{\percent} & +0.7\,\si{\percent} & +10.4\,\si{\percent} \\
            & VTM 23.10~\cite{VTM23} & +5.6\,\si{\percent} & -29.5\,\si{\percent} & -20.7\,\si{\percent} \\
            \midrule
            \multirow{6}{*}{MCL-JCV} 
            & P-SWA (ours) & -7.5\,\si{\percent} & -7.1\,\si{\percent} & -7.1\,\si{\percent} \\
            & VCT~\cite{mentzer_vct_2022} & +29.2\,\si{\percent} & +18.5\,\si{\percent} & +19.3\,\si{\percent} \\
            & DCVC-DC~\cite{li_neural_2023} & -11.7\,\si{\percent} & -40.7\,\si{\percent} & -37.3\,\si{\percent} \\
            & HM 18.0~\cite{HM18} & +37.4\,\si{\percent} & +11.8\,\si{\percent} & +18.3\,\si{\percent} \\
            & VTM 23.10~\cite{VTM23} & +7.0\,\si{\percent} & -23.1\,\si{\percent} & -16.8\,\si{\percent} \\
            \midrule
            \multirow{6}{*}{HEVC B} 
            & P-SWA (ours) & -10.0\,\si{\percent} & -5.4\,\si{\percent} & -5.8\,\si{\percent} \\
            & VCT~\cite{mentzer_vct_2022} & +22.2\,\si{\percent} & +22.8\,\si{\percent} & +22.8\,\si{\percent} \\
            & DCVC-DC~\cite{li_neural_2023} & -13.6\,\si{\percent} & -49.4\,\si{\percent} & -44.7\,\si{\percent} \\
            & HM 18.0~\cite{HM18} & +25.6\,\si{\percent} & -1.8\,\si{\percent} & +6.9\,\si{\percent} \\
            & VTM 23.10~\cite{VTM23} & -2.1\,\si{\percent} & -36.4\,\si{\percent} & -27.7\,\si{\percent} \\
            \bottomrule
        \end{tabular}
        \addtocounter{table}{-1}
    \end{minipage}
    \\[6pt] 

    \begin{minipage}[b]{0.325\textwidth}
        \centering
        \includegraphics[width=\textwidth]{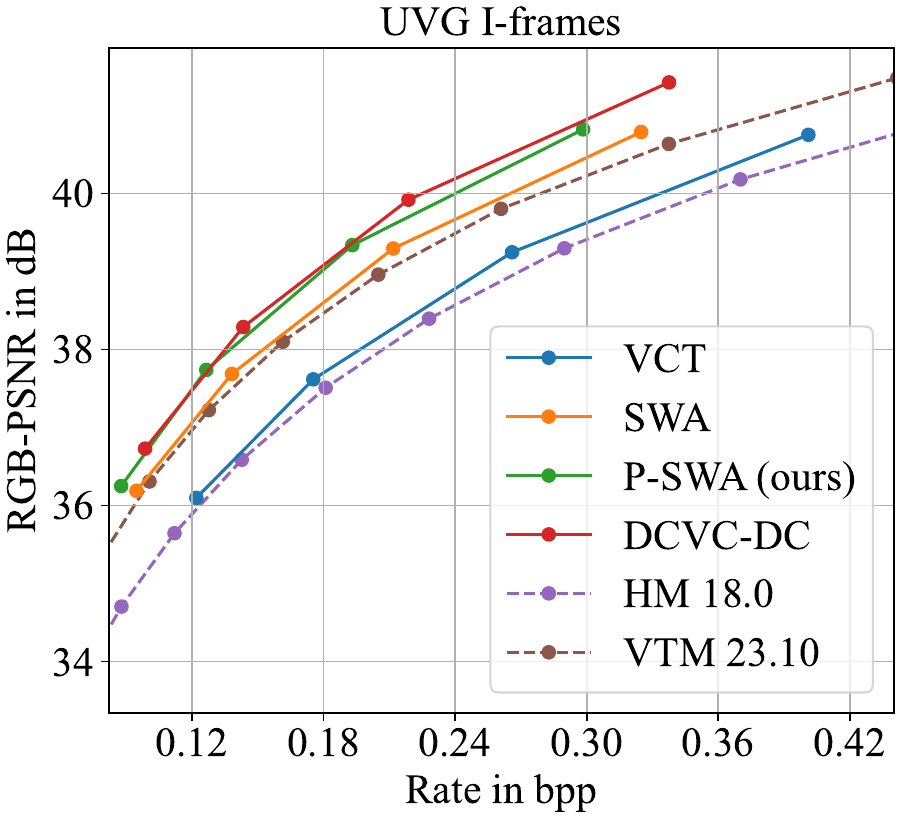}
    \end{minipage}
    \begin{minipage}[b]{0.325\textwidth}
        \centering
        \includegraphics[width=\textwidth]{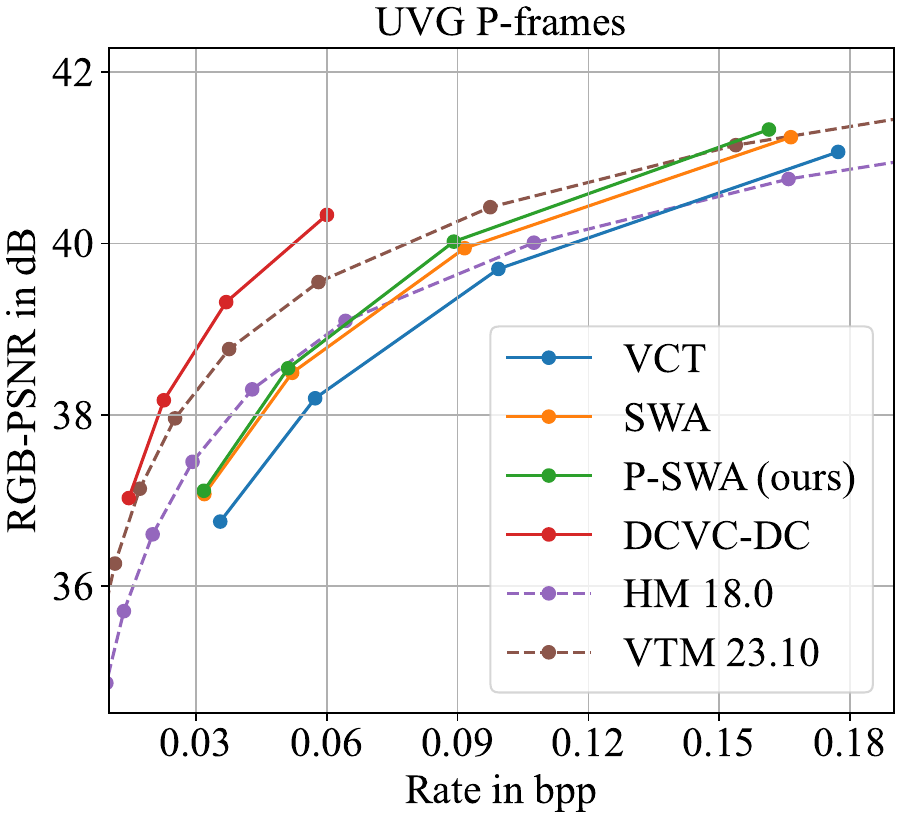}
    \end{minipage}
    \begin{minipage}[b]{0.325\textwidth}
        \centering
        \includegraphics[width=\textwidth]{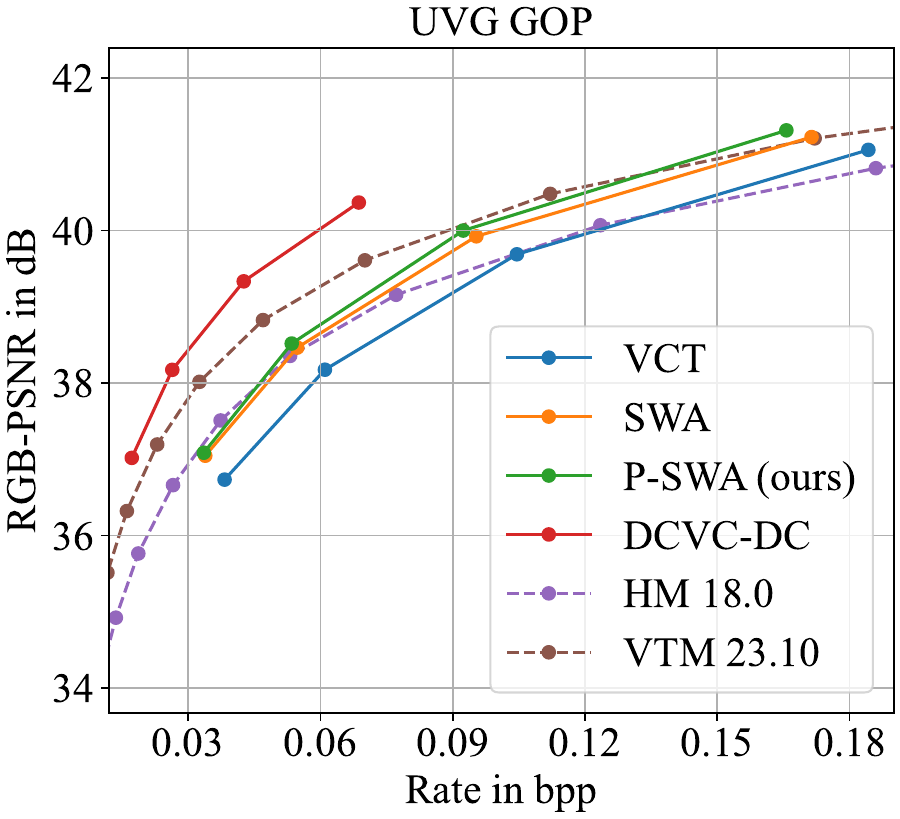}
    \end{minipage}
    \captionof{figure}{\gls{RD} curves on the UVG dataset for I-frames (\textbf{left}), P-frames (\textbf{center}), and the overall GOP (\textbf{right}).}
    \label{experiments:results:plots}
\end{figure*}

\label{sec:experiments}


\subsection{Experimental setting}


For a fair comparison, we strictly adhere to the training and testing protocols established in~\cite{kopte2025slidingwindowattentionlearned}. We utilize the feature transform from the DCVC-DC image model~\cite{li_neural_2023}, pre-trained on ImageNet~\cite{imagenet-object-localization-challenge} and subsequently frozen. Consequently, we use DCVC-DC as our anchor. The entropy model is trained in RGB space on the OpenVid-1M dataset~\cite{nan_openvid-1m_2024} using a batch size of 8. We optimize for \gls{MSE} distortion using the AdamW optimizer with default parameters. We train a single model for four rate points, using $\lambda \in \{128, 280, 680, 1600\}$. The primary training stage lasts 12 epochs on \numproduct{256 x 256} crops, utilizing every second frame to increase motion. We employ a cosine annealing learning rate schedule, decaying from \num{1e-4} to \num{1e-6}. This is followed by a two-epoch fine-tuning phase on variable resolutions up to \num{512} pixels, starting with a reduced learning rate of \num{2e-5}.

For testing, we evaluate performance on the first 96 frames of the standard UVG~\cite{uvg}, MCL-JCV~\cite{wang_mcl-jcv_2016}, and HEVC Class B~\cite{Bossen2013common} datasets with a GOP size of 32. For the learned models, source sequences are converted to RGB using the BT.709 matrix and bilinear upsampling. We also compare against \gls{HEVC}~\cite{sullivan_overview_2012} and \gls{VVC}~\cite{bross_overview_2021} anchors, specifically using the HM 18.0~\cite{HM18} and VTM 23.10~\cite{VTM23} reference models. To isolate P-frame performance, both are operated in Low Delay P configuration, excluding B-frames. They encode the intermediate YUV444 data from the upsampling step directly before converting the decoded output to RGB for PSNR calculation, which requires the RExt profile for HM. 

We configure the Temporal and Spatial Transformers with a model dimension $d=512$ and $h=16$ attention heads. Specifically, the Temporal Transformers use 8 blocks for context extraction and 4 for \gls{LRP}, while the Spatial Transformers employ 8 blocks each for the pre- and post-hyperprior stages. The Channel Transformer uses $d=1024$, $h=16$ with 2 blocks, and the convolutional hyperprior has 128 channels. To realize the parallel decoding scheme, we set the spatial wavefront step size to $s=4$ and divide the latent features into $N=4$ channel groups. In this setup, our \gls{P-SWA} context model comprises approximately \SI{136}{M} parameters, a reduction compared to the \SI{147}{M} parameters of the SWA baseline~\cite{kopte2025slidingwindowattentionlearned}. Consistent with~\cite{kopte2025slidingwindowattentionlearned}, we employ a window size of \numproduct{5 x 7 x 7} for all temporal attention mechanisms (including cross-attention) and \numproduct{7 x 7} for spatial self-attention.

\begin{table}[t]
    \centering
    \addtocounter{table}{-1}
    \vspace{-5pt}
    \caption{Comparison of encoder and decoder complexity. Precision is noted in parentheses. For DCVC-DC, complexity is averaged over a \num{32}-frame GOP, and runtime is reported for both fp32 and fp16 precision. {N/A} indicates values exceeding the measurement threshold.}
    \label{experiments:complexity:overview}
    \vspace{0.2cm}
    
    \setlength{\tabcolsep}{6pt}
    \small
    
    \sisetup{
        table-format = 4.2, 
        detect-weight,
        mode = text
    }

    \begin{tabular}{@{}l S S S[table-format=3] S[table-format=3]@{}}
        \toprule
        & \multicolumn{2}{c}{\textbf{Complexity (kMACs/px)}} & \multicolumn{2}{c}{\textbf{Runtime (ms)}} \\
        \cmidrule(lr){2-3} \cmidrule(lr){4-5}
        \textbf{Model} & {\textbf{Enc.}} & {\textbf{Dec.}} & {\textbf{Enc.}} & {\textbf{Dec.}} \\
        \midrule
        \textbf{SWA} (Mixed)        & 752.54  & 838.80  & {N/A} & {N/A} \\
        \midrule
        \textbf{P-SWA} (Mixed)      & \bfseries 624.21 & \bfseries 780.78 & \bfseries 250 & 301 \\
        \midrule
        \textbf{VCT} (Mixed)        & 2234.38 & 2320.64 & 369 & 409 \\
        \midrule
        \textbf{DCVC-DC} (fp32) & {\multirow{2}{*}{\tablenum{1307.61}}} & {\multirow{2}{*}{\tablenum{901.53}}} & 618 & 467\\
        \textbf{DCVC-DC} (fp16)     &                                   &                                      & 316 & \bfseries 273 \\
        \bottomrule
    \end{tabular}
    \addtocounter{table}{1}
\end{table}

\subsection{Results}

\subsubsection{Rate-Distortion Performance}


The \gls{BD}-rate savings relative to the \gls{SWA} baseline are presented in Table~\ref{experiments:results:table}, with corresponding rate-distortion curves for the UVG dataset shown in Fig.~\ref{experiments:results:plots}. Our proposed \gls{P-SWA} model consistently outperforms the \gls{SWA} baseline across all datasets. Notably, we achieve \gls{BD}-rate savings of up to \SI{10.0}{\percent} on I-frames and up to \SI{7.1}{\percent} for P-frames, demonstrating that our parallelized architecture not only matches but surpasses the performance of the strictly serial \gls{SWA} baseline. Consequently, \gls{P-SWA} widens the performance gap to the \gls{VCT} baseline, while offering the advantage of parallel decoding. Regarding the state-of-the-art, our method significantly narrows the performance gap to the DCVC-DC anchor for intra-coding, although a significant gap remains for inter-coding. When compared to traditional hybrid video codecs, our method consistently outperforms both HM and VTM for I-frames. For P-frames, \gls{P-SWA} surpasses the HM anchor across all tested datasets, though it does not yet match the compression efficiency of VTM.

Finally, we investigate the origin of the remaining performance gap to DCVC-DC. For I-frames, simply increasing the loss weight by a factor of 10 improves performance by \SI{4.5}{\percent} (Table~\ref{tab:ablation}), surpassing the DCVC-DC anchor. This confirms that our entropy model is not architecturally bottlenecked. We attribute the slight deficit in the standard configuration to the dominance of P-frames in the total loss, which biases the optimization, and the use of a frozen feature transform, which prevents the encoder from adapting to our new context model.

\begin{figure*}[t]
    \centering
    \begin{minipage}{0.288\textwidth}
        \centering
        \includegraphics[width=\linewidth]{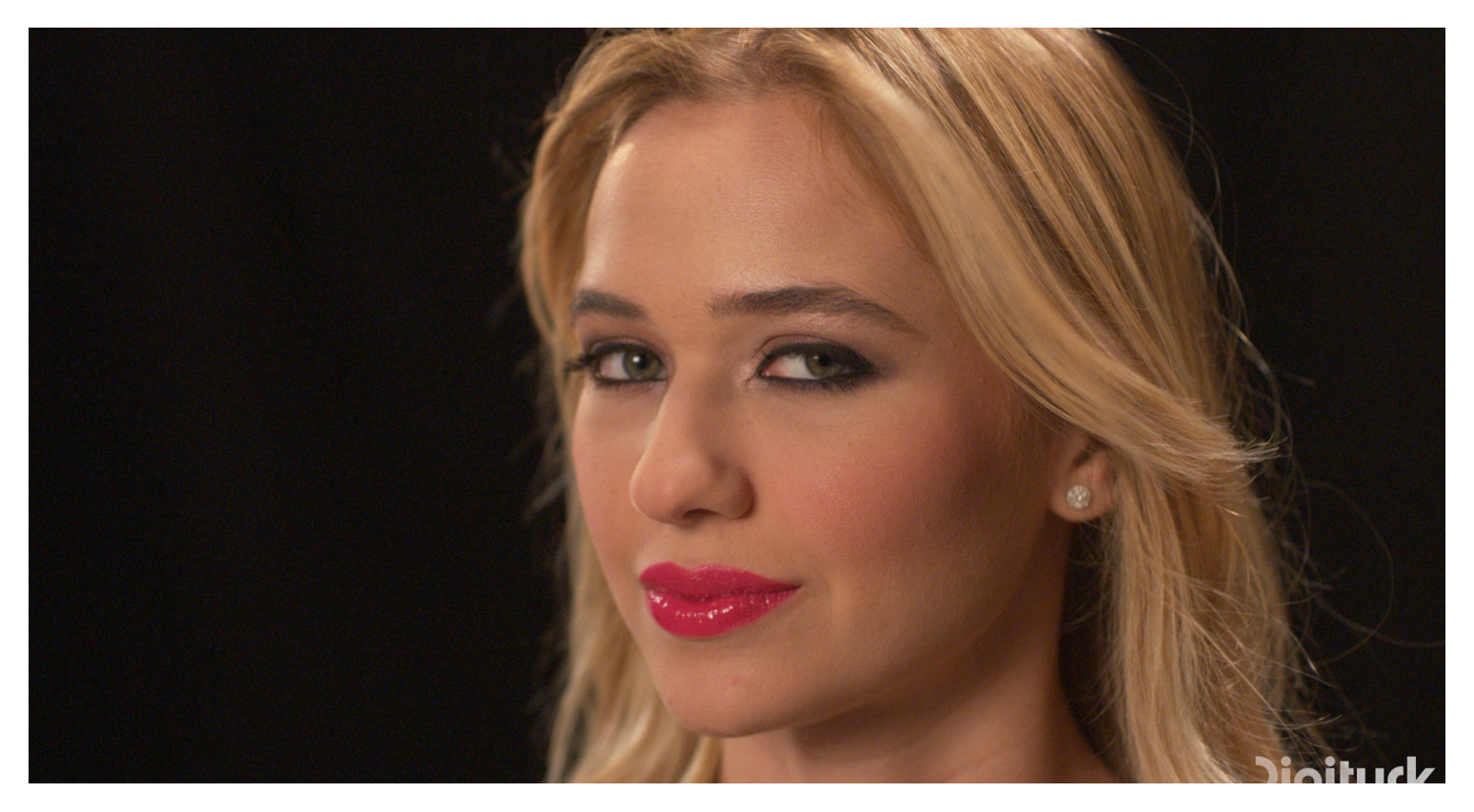}
    \end{minipage}
    \hfill
    \begin{minipage}{0.347\textwidth}
        \centering
        \includegraphics[width=\linewidth]{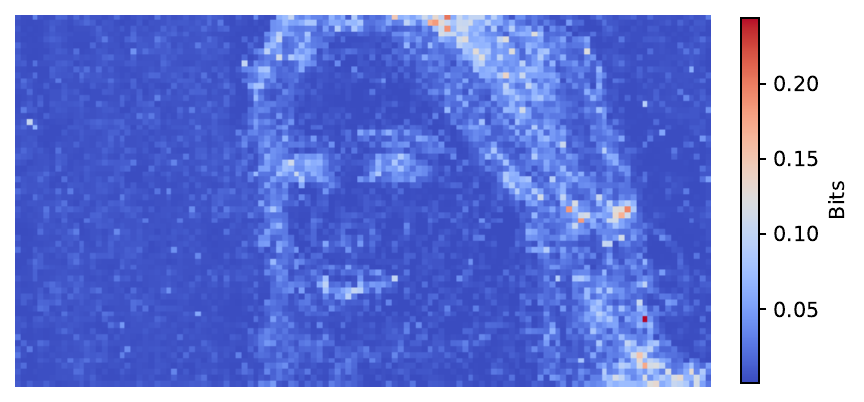}

    \end{minipage}
    \begin{minipage}{0.347\textwidth}
        \centering
        \includegraphics[width=\linewidth]{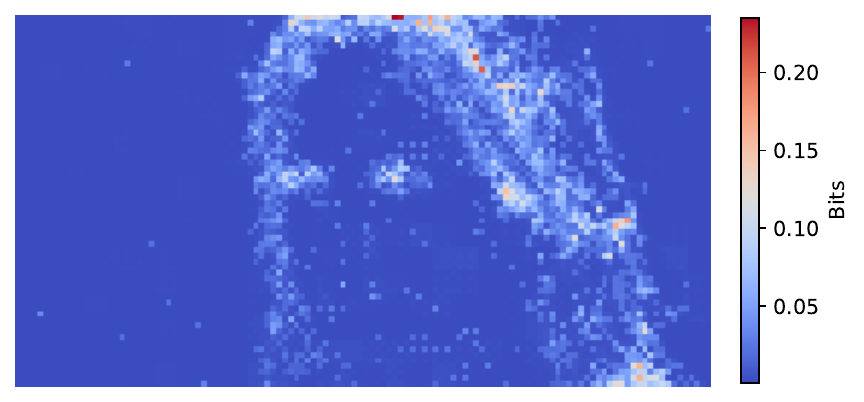}
    \end{minipage}
    
    \caption{Bit allocation comparison of the fourth frame of the UVG \textit{Beauty} sequence, encoded as a P-frame. To avoid padding artifacts, the frame is cropped to \numproduct{1920 x 1024}. \textbf{Left:} The original frame showing the static dark background and a visible watermark. \textbf{Center:} Our P-SWA model allocates unnecessary bits to the background and the static watermark due to modeling quantization noise. \textbf{Right:} DCVC-DC achieves a sparse bit distribution by effectively predicting static regions. Warmer colors indicate higher bit rate.}
    \label{fig:pframeloss}
\end{figure*}

While our model can close the gap to DCVC-DC for I-frames, a significant gap remains for P-frames. To analyze this, Fig.~\ref{fig:pframeloss} visualizes the spatial bit allocation for the fourth frame of the \textit{Beauty} sequence, which is cropped to \numproduct{1920 x 1024} for this visualization to avoid padding artifacts. The sequence features a moving subject against a static background that includes a permanent watermark. As shown in Fig.~\ref{fig:pframeloss} (right), DCVC-DC achieves extremely low rates in the static background, including on the watermark, because explicit motion compensation yields near-zero residuals. In contrast, P-SWA (center) exhibits a non-sparse bit distribution even on the completely static watermark, exposing a fundamental limitation of the \gls{VCT} paradigm. Since temporal correlation is modeled solely in the latent space, the unconditioned feature transform remains sensitive to minor input fluctuations. Consequently, noise causes feature values to fluctuate across quantization thresholds even in static areas. The entropy model must therefore model this quantization noise, incurring a rate penalty, whereas conditioning-based methods produce near-zero residuals that quantize consistently to zero, avoiding this overhead.


\subsubsection{Complexity and Runtime}
Table~\ref{experiments:complexity:overview} shows the theoretical complexity (kMACs/px) and measured runtime of our method against relevant baselines. \gls{P-SWA} demonstrates superior theoretical efficiency compared to all other baselines. However, in terms of actual runtime, our model is only \SI{36}{\percent} faster than \gls{VCT} even though \gls{VCT} has nearly three times as many MACs as our model. This discrepancy stems from memory access overheads. \gls{VCT} processes temporal context as contiguous frames, allowing for efficient, direct KV-caching. In contrast, \gls{P-SWA} utilizes advanced masking for spatial self-attention to enforce the diagonal wavefront prediction scheme, necessitating non-contiguous memory access patterns. Unlike the linear access patterns of \gls{VCT}, which benefit from high memory coalescing, our diagonal masking scheme results in scattered reads and writes during the attention computation, incurring significant latency. However, this bottleneck is not fundamental, as modern GPUs mitigate such overheads via specialized memory layouts explicitly designed for 2D spatial locality~\cite{hakura_design_1997}, as used for texture rendering.

When compared to DCVC-DC, our model has lower theoretical complexity but slightly higher decoding latency (approx. \SI{10}{\percent} slower). This comparison, however, considers a fully half-precision (fp16) implementation of DCVC-DC, whereas our \gls{P-SWA} implementation requires mixed precision, operating primarily in fp16 but utilizing fp32 within the attention mechanism for numerical stability. Furthermore, the degree of parallelization differs: DCVC-DC performs spatial and channel autoregression jointly in just 4 steps, while our decoupled design sequentially executes 4 spatial steps and 4 channel groups, totaling 16 autoregressive steps. This increased number of steps limits GPU saturation, partially negating the benefits of our lower theoretical complexity. Despite these factors, our encoder is \SI{26.4}{\percent} faster than the fp16 DCVC-DC encoder, and our decoder remains highly competitive. Beyond raw execution speed, \gls{P-SWA} also proves to be highly memory-efficient in practice, consuming less memory than the fp16 implementation of DCVC-DC (6.5 GB vs. 8.5 GB) despite its bigger model size.

\begin{table}[t]
    \centering
    \vspace{-5pt}
    \caption{BD-rate on the UVG dataset relative to our proposed P-SWA model. The top section shows the performance penalty of removing specific components. The bottom section compares our model with an intra weight of 10 against the DCVC-DC anchor.}
    \label{tab:ablation}
    \vspace{0.2cm}
    \sisetup{
        retain-explicit-plus,
        table-format = +2.1,
        detect-weight = true, 
        mode = text 
    }
    \setlength{\tabcolsep}{10.5pt} 
    \begin{tabular}{l S S}
        \toprule
        \textbf{Experiment} & \textbf{I-frames} & \textbf{P-frames} \\
        \midrule
        No Hyperprior & +12.3\,\si{\percent} & +0.1\,\si{\percent}\\
        No Channel Autoregression & +7.3\,\si{\percent} & +6.0\,\si{\percent}\\
        Simple Channel Autoregression & +2.6\,\si{\percent} & +2.6\,\si{\percent}\\
        Simple LRP & +2.1\,\si{\percent} & +13.3\,\si{\percent}\\
        \midrule
        P-SWA (Intra weight) & \bfseries -4.5\,\si{\percent} & +2.8\,\si{\percent}\\
        DCVC-DC~\cite{li_neural_2023} & -2.4\,\si{\percent} & \bfseries -47.5\,\si{\percent}\\
        \bottomrule
    \end{tabular}
\end{table}

\subsection{Ablation Study}
To validate our architectural design choices, we conduct an ablation study by removing or simplifying specific components of the P-SWA model. The results, presented in Table~\ref{tab:ablation}, show the \gls{BD}-rate increase relative to our full P-SWA model.

\textbf{Hyperprior:} Removing the hyperprior leads to a significant \SI{12.3}{\percent} degradation in I-frame performance. This is to be expected, as the initial symbols in the autoregressive sequence lack any context without this side information. However, for P-frames, the impact is negligible (\SI{0.1}{\percent}). This suggests that given temporal context, coarse spatial statistics become redundant, as the model can retrieve equivalent information from the preceding frame's latent representation.

\textbf{Channel Autoregression:} We investigate the impact of channel-wise context modeling by testing two simplified variants. First, \textit{Simple Channel Autoregression} removes the Channel Transformer and instead conditions the prediction heads on previously decoded channel groups via concatenation and masking. This results in a moderate performance drop of \SI{2.6}{\percent} across both frame types, demonstrating a small transformer captures inter-channel dependencies more effectively than simple conditioning. Second, removing channel autoregression entirely (\textit{No Channel Autoregression}) leads to a significant loss of \SI{7.3}{\percent} on I-frames and \SI{6.0}{\percent} on P-frames, highlighting the importance of modeling channel correlations.


\textbf{Latent Residual Prediction (LRP):} We replaced the temporal transformer in the \gls{LRP} module with a simple MLP prediction head (\textit{Simple LRP}), which is equivalent to the \gls{LRP} configuration in the \gls{SWA} baseline. While I-frame performance remains relatively stable, P-frame performance degrades sharply by \SI{13.3}{\percent}. We attribute this primarily to the unnecessary enforcement of spatial masking. While our proposed separate transformer remains time-causal, it operates without the restrictive prediction mask within the current frame, allowing for a more accurate prediction.

\section{Conclusion and Outlook}
\label{sec:conclusion}

In this paper, we proposed \gls{P-SWA} to overcome the inherent sequentiality of the original \gls{SWA} entropy model. By embedding the hyperprior within the transformer network and employing an accumulator to fuse its output with the previously decoded spatial context, alongside a decoupled channel autoregression mechanism, we enabled a parallel decoding scheme. Our experiments demonstrate that \gls{P-SWA} reduces decoding latency while improving coding performance. We achieve \gls{BD}-rate savings of up to \SI{10.0}{\percent} on I-frames, effectively closing the gap to the DCVC-DC anchor, and \SI{7.1}{\percent} on P-frames compared to the SWA baseline. Furthermore, the proposed method increases decoding speed by \SI{36}{\percent} compared to \gls{VCT}. 

Despite these advances, a performance gap remains for P-frames due to the susceptibility of the unconditioned latent space to quantization noise, while non-contiguous memory access patterns prevent the full translation of theoretical complexity reductions into practical speedups. Future work will therefore focus on introducing latent-space residual coding to enforce temporal consistency and optimizing memory access to better align practical runtime with theoretical efficiency.

\vfill
\newpage

\bibliographystyle{IEEEbib}
\bibliography{bib}

\end{document}